\documentclass{article}
\usepackage{amsmath} 
\newcommand{\be}{\begin{equation}}
\newcommand{\ee}{\end{equation}}

\def\bea{\begin{eqnarray}}
\def\eea{\end{eqnarray}}

\def\jrn#1#2#3#4{{#1} {\bf #2} (#4) #3}
\def\PRL{\it Phys. Rev. Lett.}
\def\PLB{\it Phys. Lett. B}
\def\PRD{\it Phys. Rev. D}
\def\NPB{\it Nucl. Phys. B}

\begin{document}

\centerline{\LARGE{\bf Seesaw and the Riddle of Mass}}
\vskip 2.5cm
\centerline{\Large P. Ramond\footnote{\uppercase{W}ork partially
supported by grant \uppercase{DE-FG}02-97\uppercase{ER}410292-4570.5 of the \uppercase{US} \uppercase{D}epartment of \uppercase{E}nergy.}}
\vskip .5cm
\centerline{Institute for Fundamental Theory, 
Department of Physics, } \vskip .2cm
\centerline{University of Florida, Gainesville FL 32611, USA 
}

\vskip 2cm

\noindent{\small{
The prediction of small neutrino masses through the Seesaw Mechanism and their subsequent measurement suggests that the natural cut-off of the Standard Model is very high indeed. The recent neutrino data must be interpreted as a reflection of physics at very high energy. We examine their implications in terms of ideas of Grand Unification and Supersymmetry, and as possible hints for a unified theory of flavor.  }}

\section{Introduction}
The Seesaw Mechanism\cite{SEESAW,MINK}, which we are here to celebrate, must be viewed in the context of the intellectual turmoil generated by the Standard Model. The renormalizability of massive Yang-Mills theories\cite{HOOFT}, the emergence  of a common description of Weak and Electromagnetic Interactions\cite{STANDARD}, and the realization that the Strong Interactions  weaken at shorter distances\cite{AF} established the Standard Model as the paradigm for all Fundamental Interactions except Gravity. Like all such paradigms, the Standard Model is (thankfully) incomplete, has suggested new puzzles of it own, and elicited many questions. None has been more dominating than Pati and Salam's\cite{PS} proposal that quarks and leptons are equal partners in one mathematical structure at very short distances, the idea of Grand-Unification. 

To appreciate the significance of the Seesaw mechanism as the link between small neutrino masses and Physics near the Planck scale, one must first describe the great theoretical speculations which led to its creation. 

\section{TRIUMPHS OF THE STANDARD MODEL}
The Fundamental Interactions (save for Gravity) are described by the Standard Model. It has withstood, practically unscathed, almost four decades of experiments, confirming {\it inter alia} its radiative structure.  All of its quarks and leptons have been discovered. Its main features are
\vskip .2cm
\noindent -- Interactions stem from three {\it weakly coupled } Yang-Mills theories based on $SU(3)$, $SU(2)$ and $U(1)$. 

\noindent --  Quarks and leptons are needed for quantum consistency: gauge anomalies cancel between quarks and leptons.

\noindent --  There are three chiral families of quarks and leptons, each with a massless neutrino.

\noindent -- The gauge symmetries are spontaneously broken: the shorter the distance, the {\it more} the symmetry.

\noindent --  It predicts a fundamental scalar particle, the Higgs boson. 
\vskip .2cm

Only one of these predictions has been proved wrong by experiments: neutrinos have masses. Today, only few of its parameters await measurement, the mass of the elusive Higgs particle, the strong CP-violating phase, and the mass of any of the three neutrinos. 

\section{ OLD \& NEW PUZZLES }
Although the successes of the Standard Model have exceeded expectations, it has a dark side:
\vskip .2cm

\noindent --  It predicts CP-violation in the Strong interaction, albeit with unknown strength.

\noindent --  It requires  Yukawa interactions without any organizing principle.

 \noindent -- It fails to {\it explain} the values of masses and mixing patterns of quarks and charged leptons. 

\noindent --  It contains too many parameters to be truly fundamental.

\noindent --  Without Gravitation it only describes the matter side of Einstein's equation, {\em sans} cosmological constant.
 
\noindent -- It fails to account for neutrino masses.
\vskip .2cm

The Standard Model presents an unfinished picture of Nature. It reminds one of the shards of a once beautifull pottery, shattered in the course of cosmological evolution.

\section{ GRAND UNIFICATION}
The quantum numbers of the three chiral families of quarks and leptons strongly suggest a more unified picture. 
 Pati and Salam's original idea is, remarkably enough, realized by unifying the three gauge groups of the Standard Model into one. In the simplest, $SU(5)$\cite{GG}, each family appears in two representations. In $SO(10)$\cite{FM}, they are grouped in its fundamental spinor representation, by adding a right-handed neutrino for each family. At the next level, we find $E_6$\cite{ESIX} where each family contains several right-handed neutrinos as well as vector-like matter. Organizing the elementary particles into these beautiful structures    
\vskip .2cm

 \noindent -- Unifies the three gauge groups.

\noindent -- Relates Quarks and Leptons.

\noindent -- Explains anomaly cancellations.
\vskip .2cm

There are indications that this idea ``wants to work". When last seen, the three coupling constants of the Standard Model are perturbative. Using the renormalization group equations to continue them  deep into the ultraviolet, they  get closer to one another, but fail to meet at one scale: the quantum number patterns did not quite match the dynamical information. This near (thought at the time to be exact) unification introduced Planck scale physics into the realm of particle physics.

One by-product of Grand Unification is violation of baryon number. Hitherto unobserved, proton decay remains one of the most important consequences 
from these ideas. In a serendipitous twist, proton decay detectors now serve as the telescopes of neutrino astronomy! Other global symmetries also bite the dust: the relative lepton numbers are violated in $SU(5)$ and $SO(10)$ violated the total lepton number as well, and the extraordinary limits on these processes are consistent with  the grand-unified scale.

\section{GRAND-UNIFIED LEGACIES}
Grand Unification by itself does not yet have any direct experimental vindication; it is an incubator of new ideas that, even today, drive speculations on the Physics at extra-short distances. 
\vskip .2cm

\noindent --  It linked the large grand-unified scale to tiny neutrino masses\cite{SEESAW}.

\noindent -- It suggested relations between quark and charged lepton masses, although the flavor riddles of the Standard Model remain unexplained.

\noindent -- It  created the ``gauge hierarchy" problem: why quantum corrections keep the ratio of the Higgs mass to the Unification scale small. 
\vskip .2cm

Moreover, two of its predictions have linked particle physics to pre-Nucleosynthesis Cosmology:
\vskip .2cm

\noindent -- The possibility  of monopoles in our universe led to the idea of Inflationary Cosmology\cite{GUTH}, which solves many long standing puzzles and whose prediction of a flat universe has been recently verified. 

\noindent -- Proton decay. This offered a framework for understanding the baryon asymmetry\cite{YOSHIMURA} of the Universe.   
\vskip .2cm

Today, only one of these predictions, tiny neutrino masses, has been borne out by experiment. On the conceptual side, it has also provided an alternative mechanism for the generation of Baryon asymmetry of the Universe through a primordial lepton asymmetry\cite{LEPTOGENESIS}. Still, Grand Unification is at most a partial theory of Nature, since it does not address Gravity (space-time is either flat or a fixed background ), nor the origin of the three chiral families and its associated flavor puzzles.

\section{ SUPERSTRINGS}
 At the 1973 London conference, David Olive declared Superstring Theories to be candidate ``Theories of Everything", since they reproduce Einstein's gravity at large distances with no ultraviolet divergences, and  also contain (some) gauge theories. This view has since gained much credence and notoriety. The  matter content has gotten much closer to reality\cite{HET}, although this unification of the gravitational and gauge forces takes place in a somewhat unsettling background:
\vskip .2cm
\noindent -- Fermions and Bosons are related by a new type of symmetry: {\it Supersymmetry}\cite{PMR}.      

\noindent -- Ultimate Unification takes place in nine or ten space dimensions!
\vskip .2cm

Nature at the millifermi displays neither Supersymmetry nor extra space dimensions. Yet, the lesson of the Standard Model of more symmetries at shorter distances provide an argument for these to be fabrics of the Ultimate Theory; these symmetries are somehow destroyed in the process of cosmological evolution.  To compare the highly symmetric superstring theories to Nature, a dynamical understanding of their breakdown is required, an understanding that  still eludes us. 

To relate to Nature, experiments at energies at which these symmetries appear must be carried out. All could be just around the energy corner, although   circumstantial evidence lends more credence to low- energy Supersymmetry than to low-energy extra dimensions. The collapse of the extra space dimensions occurs first, while Supersymmetry hangs on to later times (lower energies). It is a challenge to theory to find a dynamical reason which triggers the breakdown of higher-dimensional space (perhaps through  brane formation), while leaving Supersymmetry nearly intact.


\section{ SUPERSYMMETRY}
 Supersymmetry is  an attractive theoretical concept; it is required by the unification of gravity and gauge interactions, and links fermions and bosons. Also, the mass of the spinless superpartner of a Weyl fermion, inherits quantum-naturality\cite{GILDENER} through  the chiral symmetry of its partner.  
 
Morever, when applied to the Standard Model, it yields quantitative predictions that fit remarkably well with Gauge Unification. With Supersymmetry,
\vskip .2cm

\noindent --   The Gauge hierarchy problem is managed: the Higgs mass is stabilized even in the presence of a large (grand-unification) scale

\noindent --  The three gauge couplings of the Standard Model run to a single value in the deep ultraviolet with the addition of superpartners in the TeV range. Thus naturally emerges a new scale  using the renormalization group, a scale that matches the quantum number patterns of the elementary particles. 

\noindent -- With supersymmetry the renormalization group displays an infrared fixed point that predicts\cite{PENDLETON} the top quark mass, in agreement with experiment.

\noindent -- Under a  large class of ultraviolet initial conditions, the same renormalization group shows that the breaking of supersymmetry triggers electroweak breaking\cite{EWBREAK}. 
\vskip .2cm

Supersymmetry at low energy is the leading theory for physics beyond the Standard Model, although  many puzzles remain unanswered and  new ones are created as well. 

For one, there are almost as many theories of  supersymmetry breaking as there are theorists, and  none, theories and theorists alike, are convincing. It is an experimental question.

In addition, Supersymmetry deepens the flavor riddles of the Standard Model by  predicting  new scalar particles which generically produce flavor-changing neutral processes. Even if the breaking mechanism is flavor-blind (tasteless), non-trivial effects are expected: supersymmetry-breaking is already highly constrained by the existing data set.

The existence of low-energy Supersymmetry will soon be tested at the LHC.  May the supersymmetry-breaking mechanism parameters prove to be so unique as to allow intellectually-challenged theorists (the author included) to infer its origin from the LHC data alone!

\section{ MINUTE NEUTRINO MASSES}
 The only solid experimental evidence to date for physics beyond the Standard Model is the observation of oscillation among neutrino species.  Thirty five years of experiments on solar neutrinos,  Homestake\cite{Homestake}, GALLEX\cite{GALLEX}, SAGE\cite{SAGE}, SUPERK\cite{SKsol} and SNO\cite{SNO}, yield 

 $$\Delta m^2_\odot~=~\vert~m^2_{\nu_1}-m^2_{\nu_2}~\vert~\sim~7.\times 10^{-5}_{}~{\rm eV^2}\ ,$$
with corroborating evidence on antineutrinos\cite{Kamland}. Neutrinos born in Cosmic ray collisions\cite{SKatm}, and on earth\cite{K2K} give
     
$$\Delta m^2_\oplus~=~\vert~m^2_{\nu_2}-m^2_{\nu_3}~\vert~\sim~3.\times 10^{-3}_{}~{\rm eV^2}\ .$$
The best bound to their absolute value of the masses comes from WMAP\cite{WMAP}

$$\sum_i~m^{}_{\nu_i}~<~.71~{\rm eV}\ .$$
These experimental findings are not sufficient to determine fully the mass patterns. One oscillates between three patterns, {\it hierarchy}, 

$$|m_{\nu _1}| < |m_{\nu _2}| \ll |m_{\nu _3}|\ ,$$
{\it inverse hierarchy}

$$|m_{\nu _1}| 
\simeq |m_{\nu _2}| \gg |m_{\nu _3}|\ ,$$
and {\it hyperfine}

$$|m_{\nu _1}| \simeq |m_{\nu _2}| \simeq |m_{\nu _3}|\ .$$
The mixing patterns provide some surprises, since it contains one small angle and two large angles. In terms of the MNS mixing matrix,

$${\begin{pmatrix}\cos\theta^{}_\odot&\sin\theta^{}_\odot&\epsilon\cr
-\cos\theta^{}_\oplus~\sin\theta^{}_\odot&\cos\theta^{}_\oplus~\cos\theta^{}_\odot&\sin\theta^{}_\oplus\cr
\sin\theta^{}_\oplus~\sin\theta^{}_\odot&-\sin\theta^{}_\oplus~\cos\theta^{}_\odot&\cos\theta^{}_\oplus\end{pmatrix}}\ ,$$
the various experiments yield
    
$$ \sin^2 2\theta^{}_\oplus~>~0.85\ ,\qquad  0.30~<~ \tan^2\theta^{}_\odot~< ~0.65 \ ,$$
while there is a only a limit\cite{CHOOZ} on the third angle 

$$ \vert~\epsilon~\vert^2_{}~<~0.05\ .$$
Spectacular as they are, these results generate new questions for experimenters:
\vskip .2cm
\begin{itemize}
\item Are the masses Majorana-like (i.e. lepton number violating)?

\item What are their absolute values? 

\item Can one measure the sign of $\Delta m^2$?

\item What is the value of the CHOOZ angle?

\item Is  CP-violation in the lepton sector observable?

\end{itemize}

\noindent They also generate new theoretical questions

\begin{itemize}

\item  Are there right-handed neutrinos? 

\item  How many? How heavy, and with what hierarchy? 

\item   Where do they live? Brane or bulk? 

\item  Do their decays trigger leptogenesis\cite{LEPTOGENESIS}?
\end{itemize}

\section{ Standard Model Analysis}
In the context of Grand Unification, one needs to discuss both quark and lepton mass matrices. To that effect, recall that the masses and mixings of the quarks are determined from the diagonalization of Yukawa matrices generated by the  $\Delta I_{\rm W}=\frac{1}{2}$ breaking 
of electroweak symmetry, for charge $2/3$ 

$$
{\mathcal U}^{}_{2/3}\,
{\begin{pmatrix}m^{}_u&0&0\cr 0&m^{}_c&0\cr 0&0&m_t^{}\end{pmatrix}}
\,{\mathcal V}^{\dagger}_{2/3}\ ,$$
and charge $-1/3$

$$
{\mathcal U}^{}_{-1/3}\,
{\begin{pmatrix}m^{}_d&0&0\cr 0&m^{}_s&0\cr 0&0&m_b^{}\end{pmatrix}}
\,{\mathcal V}^{\dagger}_{-1/3}\ ,
$$
resulting in the observable CKM matrix

$$
{\mathcal U}^{}_{CKM}~\equiv~{\mathcal U}^{\dagger}_{2/3}\,{\mathcal U}^{}_{-1/3}\ ,
$$
which, up to corrections of the order of the Cabibbo angle, $\theta_c\sim 13^\circ$,  is equal to the unit matrix. This implies similar family mixings 
for up-like and down-like quarks. Their masses are of course highly hierarchical. The charged lepton Yukawa matrix  

$${\mathcal U}^{}_{-1}\,
{\begin{pmatrix}m^{}_e&0&0\cr 0&m^{}_\mu&0\cr 0&0&m_\tau^{}\end{pmatrix}}
\,{\mathcal V}^{\dagger}_{-1}$$
also stems from $\Delta I_{\rm W}=\frac{1}{2}$ electroweak breaking, and has hierarchical eigenvalues. 

To obtain neutrino masses in the Standard Model, it is simplest to add one right-handed neutrino for each family. This yields another $\Delta I_{\rm W}=\frac{1}{2}$ Yukawa matrix

$${\mathcal U}^{}_{0}\,{\begin{pmatrix}m^{}_1&0&0\cr 0&m^{}_2&0\cr 0&0&m_3^{}\end{pmatrix}}\,{\mathcal V}^{\dagger}_{0}\ ,$$
but does not explain the extraordinary gap between charged and neutral leptons. 

The right-handed neutrino masses are of Majorana type, since they have no gauge quantum numbers to forbid it (unlike electrons, say), and  necessarily violate total lepton number.

In the context of effective field theories, one expects their masses to be of the order of lepton number breaking. Total lepton number-violating processes have never been seen resulting in a bound from neutrinoless double $\beta$ decay experiments. So either they are very large or zero. 

If they are zero, the analysis proceeds as in the quark sector, and the observable MNS lepton mixing matrix is just

$${\mathcal U}^{}_{MNS}~\equiv~ {\mathcal U}^{\dagger}_{-1}\,{\mathcal U}^{}_{0}\ .$$
As for the quarks, it would be generated solely from the isospinor breaking of electroweak symmetry, even though the mixing patterns are so different. 

In the belief that global symmetries are an endangered species (for one, black holes eat them up), we expect their masses to set the scale of the Standard model's cut-off, since they are unprotected by gauge symmetries. This yields the seesaw where large right-handed masses engender tiny neutrino masses, the latter being suppressed over that of the charged particles by the ratio of the two scales
     
$$\frac{\Delta I_{\rm W}=\frac{1}{2}}{\Delta I_{\rm W}=0}\ ,$$
thus  introducing a large electroweak-singlet scale in the Standard Model. The neutrino mass matrix is then  
 
$${\mathcal M}^{(0)}_{Seesaw}~=~{\mathcal M}^{(0)}_{  {Dirac}}\,
\frac{1}{{\mathcal M}^{(0)}_{   {Majorana}}}\,{\mathcal M}^{(0)\,T}_{  {Dirac}}\ ,$$
which we can rewrite as

$${\mathcal M}^{(0)}_{   {Seesaw}}~=~{\mathcal U}^{}_{0}\,\,
  {\bf{\mathcal C}}\,\,{\mathcal U}^{T}_{0}\ ,$$
in terms of the central matrix\cite{DLR}

$$ {\mathcal C}~=~{\mathcal D}_0^{}\,{\mathcal V}^{\dagger}_{0}\,\frac{1}{{\mathcal M}^{(0)}_{   {Majorana}}}\,
{\mathcal V}^{*}_{0}\,{\mathcal D}_0^{}\ .$$
It is diagonalized by the unitary matrix ${\mathcal F}$
 
$$~~~  {\mathcal C}~=~   {\mathcal F}\,{\mathcal D}^{}_\nu\,   {\mathcal F^{\,T}_{}}\ ,$$
where the mass eigenstates produced in $\beta$-decay are (unimaginatively labelled  as ``1", ``2", ``3")      

$$
{\mathcal D}_\nu^{}~=~{\begin{pmatrix}m^{}_{\nu_1}&0&0\cr 0&m^{}_{\nu_2}&0\cr 0&0&m_{\nu_3}^{}\end{pmatrix}}\ .$$
The effect of the Seesaw  is to add the unitary ${\mathcal F}$ matrix to the MNS lepton matrix 

$${\mathcal U}^{}_{MNS}~=~ {\mathcal U}^{\dagger}_{-1}\,{\mathcal U}^{}_{0}\,\,   {\mathcal F}\ .$$
This framework enables us to recast  theoretical questions in terms of $\mathcal F$. In particular, where do the large angles come from? We  
catalog models in terms of the number of large angles contained in $\mathcal F$, none, one or two? 

\section{A Modicum of Grand Unification}  
To answer that question, we must turn to Grand Unification ideas for guidance, where relations between the $\Delta I^{}_{\rm W}=\frac{1}{2}$ quark and lepton Yukawa matrices appear naturally. 

In $SU(5)$, the charge $-1/3$ and charge $-1$ Yukawa matrices are family-transposes of one another. 

$$ {\mathcal M}^{(-1/3)}_{}~\sim~{\mathcal M}^{(-1)\,T}_{}\ .$$
In $SO(10)$, it is the charge $2/3$ Yukawa matrix that is related to the Dirac charge $0$ matrix

 $$ {\mathcal M}^{(2/3)}_{}~\sim~{\mathcal M}^{(0)}_{  {Dirac}}\ .$$
These result in naive expectations for the unitary matrices that yield observable mixings

$${\mathcal U}^{}_{-1/3}~\sim~{\mathcal V}^{*}_{-1}\ ;\qquad {\mathcal U}^{}_{2/3}~\sim~{\mathcal U}^{}_{0}\ .$$
Assuming this pinch of grand-unification, we can relate the CKM and MNS matrices

\bea
\nonumber{\mathcal U}^{}_{MNS}&=& {\mathcal U}^{\dagger}_{-1}\,{\mathcal U}^{}_{0}\,   {\mathcal F}\cr 
 \nonumber&\sim&{\mathcal U}^{\dagger}_{-1}\,{\mathcal U}^{}_{-1/3}\,{\mathcal U}^{\dagger}_{CKM}\,   {\mathcal F} \cr
&\sim&     \Big({{\mathcal V}^T_{-1/3}\,{\mathcal U}^{}_{-1/3}}\,\Big)\,{\mathcal U}^{\dagger}_{CKM}\,\,{\mathcal F}\eea
  Hence two wide classes of models:
\vskip .2cm
\noindent I-)  Family-Symmetric ${\mathcal M}^{}_{-1/3}$ Yukawa matrices. 
In these we have  

$${{\mathcal U}^{}_{-1/3}}~=~{{\mathcal V}^*_{-1/3}}\ ,$$
so that      

$$
{{{\mathcal U}^{}_{MNS}~=~ {\mathcal U}^{\dagger}_{CKM}\,\,   {\mathcal F}
}}\ .$$
In these models, ${\mathcal F}$ necessarily contains two large angles. In the absence of any symmetry acting on $\mathcal F$, these models require a highly structured $\mathcal F$ matrix, which could even be non-Abelian.

Interestingly, these  models provide a testable prediction for the size of the CHOOZ angle. With a family-symmetric charge $-1/3$ matrix, the MNS matrix reads

\bea
& &{\mathcal U}^{}_{MNS}~=~{\mathcal U}^{\dagger}_{CKM}\,\times \cr & &\cr
& &{\begin{pmatrix}\cos\theta^{}_\odot&\sin\theta^{}_\odot&   {\lambda^\gamma}\cr
-\cos\theta^{}_\oplus~\sin\theta^{}_\odot&\cos\theta^{}_\oplus~\cos\theta^{}_\odot&\sin\theta^{}_\oplus\cr
\sin\theta^{}_\oplus~\sin\theta^{}_\odot&-\sin\theta^{}_\oplus~\cos\theta^{}_\odot&\cos\theta^{}_\oplus\end{pmatrix}}
\ ,\nonumber\eea
where we have chosen to fill the zero in the $\mathcal F$ matrix by a Cabibbo effect, with $\gamma$ presumably greater than one. It follows that 

$$
\theta^{}_{13}~\sim~\lambda\sin\theta_\oplus~\sim~\frac{1}{\sqrt 2}\,\lambda\ .
$$
It will be interesting to see if this definite prediction of type I models, $\theta_{13}\sim 7-9^\circ$, is borne out in future experiments.
\vskip .2cm

\noindent II-)  Family-Skewed ${\mathcal M}^{}_{-1/3}$ Yukawa matrices. One can make a compelling arguments for at least one large angle to reside in $\mathcal U_{-1}$. If we extend the Wolfenstein\cite{WOLF} expansion of the CKM matrix in powers of the Cabibbo angle $\lambda$ to include quark mass ratios

$$\frac{m^{}_s}{m^{}_b}~\sim~{\lambda^2_{}}\qquad \frac{m^{}_d}{m^{}_b}~\sim~{\lambda^4_{}}\ ,$$
we find the charge $-1/3$ Yukawa matrix
 
$${\mathcal M}^{(-1/3)}_{}~=~  {\begin{pmatrix}   \lambda^4_{}&   \lambda^3_{}&   \lambda^3_{}\cr
    \lambda^?_{}&   \lambda^2_{}&   \lambda^2_{}\cr
    \lambda^?_{}&    \lambda^{?}_{}& 1\end{pmatrix}}\ .$$
If the exponents are related to charges, as in the Froggatt-Nielsen\cite{FN} schemes, the lower diagonal exponents are known, and we get the orders of magnitude
 
$${{\mathcal M}^{(-1/3)}_{}}~=~  
{\begin{pmatrix}   {\lambda^4_{}}&   {\lambda^3_{}}&   {\lambda^3_{}}\cr
    {\lambda^3_{}}&   {\lambda^2_{}}&   {\lambda^2_{}}\cr
    {\lambda^1_{}}&    1&1\end{pmatrix}}\ ,$$
which is not  family-symmetric. In the limit of no Cabibbo mixing, 

$${\mathcal M}^{(-1/3)}_{}~\approx~{\begin{pmatrix}0&0&0\cr 0&0&0\cr 0&  {a}&  b\end{pmatrix}}
+{\mathcal O}({   \lambda})\ ,$$
 and 

$${\mathcal U}^{}_{MNS}~=~
{\begin{pmatrix}1&0&0\cr 0&\cos\theta^{}_\oplus&\sin\theta^{}_\oplus\cr 0&-\sin\theta^{}_\oplus&\cos\theta^{}_\oplus\end{pmatrix}}\,   {\mathcal F}\ ,$$
where 
     
$$\tan\theta^{}_\oplus~=~\frac{a}{b}\ , $$
is of order one\cite{ILR}. In these models, ${\mathcal F}$ need contain only one large angle, which is very natural, although they give no generic prediction for the CHOOZ angle.

\section{Right-Handed Hierarchy}
In most models, $\mathcal F$ must contain at least one large angle to accomodate the data. This presents a puzzle since $\mathcal F$ diagonalizes a matrix which contains the neutral Dirac Yukawa matrix which is presumably hierarchical, coming from the isospinor electroweak breaking. 
This suggests special restrictions put upon the Majorana mass matrix of the right-handed neutrinos. We want to illustrate this point by looking at a $2\times 2$ two-families case\cite{DLR}, and write

 $$
{\mathcal D}^{}_0~=~m{\begin{pmatrix}a\,{\lambda^\beta_{}}&0\cr 0&1\end{pmatrix}}\ , $$
and define $M^{}_1\ , M^{}_2$ to be the eigenvalues of the right-handed neutrino's Majorana mass matrix. This matrix can be diagonalized by a large mixing angle in one of two cases:
\vskip .2cm

\noindent -- Its matrix elements have similar orders of magnitude ${\mathcal C}_{11} ~\sim~{{\mathcal C}_{22}} ~\sim~{\mathcal C}_{12}$, in which case we find that 

$$
\frac{M_1}{M_2}~\sim~   {\lambda^{2\beta}_{}}\ ,$$
suggesting a doubly {\it correlated hierarchy} betwen the $\Delta I_{\rm W}=0$ and $\Delta I_{\rm W}=\frac{1}{2}$ Sectors. This agrees well with grand-unified models such as $SO(10)$ and $E_6$, where each  right-handed neutrinos is part of a family.
\vskip .2cm

\noindent --A large mixing angle can occur if the diagonal elements are much smaller than the diagonal ones, that is 
${\mathcal C}_{11}\, , \,{\mathcal C}_{22} ~\ll~ {\mathcal C}_{12}$. Then we find 

$$\frac{\lambda^\alpha~m^2}{\sqrt{-M_1 M_2}}\,
{\begin{pmatrix}0&a\cr a&0\end{pmatrix}}\ .$$
Hence maximal mixing may infer that some of the right-handed neutrinos are Dirac partners of one another, leading to conservation in the right-handed mass matrix, of a relative lepton number $L_1-L_2$. 

\section{Cabibbo Flop}
As we have seen, Grand-Unification, even in its simplest form, implies Cabibbo-sized effects in the MNS matrix. In the quark sector, Cabibbo mixing is the strongest between the first and second family. Applied to the lepton sector, the solar angle may be maximal, with a Cabibbo correction of $13^\circ$\cite{FUJIHARA,SMIRNOV}. 

Recently, we\cite{DER} have been exploring possible Wolfenstein parametrizations of the MNS matrix, in the hope that some regularity might emerge from the data, once Cabibbo effects are taken into account. 

Since we do not know how the Cabibbo angle is generated in flavor theories. So we start by asking if the limit ${\theta_c^{}\rightarrow 0}$ makes any theoretical sense. To simplify matters, assume there is only one small parameter in the flavor sector; then the quark and charged lepton masses of the first two families are zero. In the same limit, ${\mathcal U}_{CKM}=1$, and there are no neutral flavor changes. Of course the mixing between the first two families is undetermined.

We do not know ${\mathcal U}_{MNS}$ in that limit, the starting point of a Wolfenstein parametrization for the lepton mixing matrix. 

The measured values of the lepton mixing angles are

$$\theta_\odot^{}={32.5^\circ_{}}^{\,+\,2.4^\circ}_{ \,-\,2.3^\circ} \ ;\quad \theta^{}_\oplus~=~ {45.00^\circ_{}}^{\,+10^\circ}_{\,-10^\circ} \ ;\quad \theta_{CHOOZ}< 13^\circ_{}\ .$$
The solar angle is well measured, but the atmospheric angle is not, and could very well be non-maximal. Furthermore, their values could be affected by Cabibbo flop of $\pm ~13^\circ_{}$, and the CHOOZ angle could well be a Cabibbo effect. Our starting point is 
 
$${\mathcal U}^{}_{MNS}= {\begin{pmatrix}\cos{\eta_\odot} &\sin{\eta_\odot} &0\cr -\cos{\eta_\oplus}\sin{\eta_\odot}&\cos{\eta_\oplus}\cos{\eta_\odot}
&\sin{\eta_\oplus}\cr
\sin{\eta_\oplus}\sin{\eta_\odot}&-\sin{\eta_\oplus}\cos{\eta_\odot}&\cos{\eta_\oplus}\end{pmatrix}}~+~\cdots\ ,$$
with a range of initial angles 

$$15^\circ_{}~<~ \eta^{}_\odot~<~ 45^\circ_{}\ ;\qquad 30^\circ_{}~<~ \eta^{}_\oplus~<~ 60^\circ_{}\ .$$
We write the Wolfenstein expansion of the MNS Matrix in the form 

$${\mathcal U}^{}_{MNS}~\equiv ~{\mathcal W}~+~{{\mathcal O}(\lambda)}\ ,$$
where the starting matrix is split in two parts, showing the large angles

$${\mathcal W}~=~{\mathcal W}_\oplus\,{\mathcal W}_\odot \ ,$$
 with

$${\mathcal W}_\oplus~=~{\begin{pmatrix}1&0&0\cr 0&\cos{\eta_\oplus}&- \sin{\eta_\oplus}\cr 0&\sin{\eta_\oplus}&\cos{\eta_\oplus}\end{pmatrix}}\ ,$$

$${\mathcal W}_\odot~=~{\begin{pmatrix}\cos{\eta_\odot} & \sin{\eta_\odot} &0\cr -\sin{\eta_\odot} &\cos{\eta_\odot} &0\cr 0&0&1\end{pmatrix}}\ .$$
We introduce Cabibbo flop through the unitary matrix

$${\mathcal V}~=~I+\Delta(\lambda)\ ,$$
with ${\Delta(0)~=~1}$. Unlike the quark sector it does not commute with the starting matrix

$$ [\,{\mathcal W}\,,\,{{\mathcal V}(\lambda)}\,]~\neq~0\ .$$
This means that Cabibbo effects from the left and from the right or even in between the two starting matrices are not equivalent. Hence we consider basic flops

 \begin{itemize}
\item  {{ Left }} ~~~~~\,${\mathcal U}_{MNS}={{\mathcal V}(\lambda)}\,{\mathcal W}\,$
\item  {{ Right }}~~~\, ${\mathcal U}_{MNS}={\mathcal W}\,{{\mathcal V}(\lambda)}$
\item  {{ Middle }}~~ ${\mathcal U}_{MNS}={\mathcal W}_\oplus{{\mathcal V}(\lambda)}\,{\mathcal W}_\odot\,$
\end{itemize}
and we can have one ${\mathcal O}(\lambda)$ correction (single flop), or two (double flop). The present data is not sufficient to single out a particular 
 Wolfenstein parametrization, but the hope is that by considering possible Cabibbo effects on various starting matrices, generic features suggestive of flavor patterns might become obvious. In particular, they would restrict the size of the CHOOZ angle and of the  CP-violation.

To illustrate these points, consider the effect of flop matrices, shown here to ${\mathcal O}(\lambda^3)$,   

$${\mathcal V}^{}_{12}(\lambda)~=~{\begin{pmatrix}   1-\frac{a^2}{2}\lambda^2&a\,\lambda&b\,\lambda^2\cr -a\,\lambda&1-\frac{a^2}{2}\lambda^2&0\cr -b\,\lambda^2&0&1\end{pmatrix}}$$

$${\mathcal V}^{}_{23}(\lambda)~=~{\begin{pmatrix}   1&0&b\,\lambda^2\cr 0&1-\frac{a^2}{2}\lambda^2&a\,\lambda\cr -b\,\lambda^2&-a\,\lambda&1-\frac{a^2}{2}\lambda^2\end{pmatrix}}$$

$${\mathcal V}_{\rm double}(\lambda)~=~{\begin{pmatrix}   1-\frac{a^2}{2}\lambda^2&a\,\lambda&(b+\frac{aa'}{2})\,\lambda^2\cr -a\,\lambda&1-\frac{a^2+a'^2}{2}\lambda^2&a'\,\lambda\cr (\frac{aa'}{2}-b)\,\lambda^2&-a'\,\lambda&1-\frac{a'^2}{2}\lambda^2\end{pmatrix}}\ ,$$
where we have limited ourselves to $a=\pm 1\ ;\quad a'=\pm 1\ ;\quad 0.8<~b~<~ 1.2$. For instance, a left single flop ${\mathcal V}_{23}$, yields values for the starting angles that are different from the data, 
\vskip .5cm
\begin{center}
\begin{tabular}{|c|c|c|c|c|}
\hline
$~{\eta^\circ_\odot} ~$& $~{\eta^\circ_\oplus} ~$&${\theta^\circ_\odot}$&${\theta^\circ_\oplus}$ &${\theta^\circ_{13}}$ \\
 \hline \hline         
$~ 30~ $&$ ~30~$  &$\sim 31~$& $43$&$.06-.4$\\
 \hline    
$~30~$& $~60~$  &$31-32$&$\sim 47$&$.6-2.5$\\
\hline
 \end{tabular}\end{center}
\vskip .5cm
Right single flops with ${\mathcal V}_{23}$ and ${\mathcal V}_{12}$ produce: 
 \vskip .5cm
\begin{center}
\begin{tabular}{|c|c|c|c|c|}
\hline
$~{\eta^\circ_\odot} ~$& $~{\eta^\circ_\oplus} ~$&${\theta^\circ_\odot}$&${\theta^\circ_\oplus}$ &${\theta^\circ_{13}}$ \\
 \hline \hline          
$~30~$& $~60~$  &$30-31$&$ 48-50$&$3-10$\\
\hline
$~15~$& $~45~$& $\sim 30.3$ &$44-45$&$2-4$\\                                                            
 \hline
$~45~$&$~45~$&$~\sim 32$&$44-46$&$1-3$\\ 
\hline
 \end{tabular}\end{center}
\vskip .5cm
A right double flop with ${\mathcal V}_{\rm double}$:
 \vskip .5cm
\begin{center}
\begin{tabular}{|c|c|c|c|c|}
\hline
$~{\eta^\circ_\odot} ~$& $~{\eta^\circ_\oplus} ~$&${\theta^\circ_\odot}$&${\theta^\circ_\oplus}$ &${\theta^\circ_{13}}$ \\
 \hline \hline          
$~15~$& $~60~$  &$ \sim 30.3$&$ 45-51$&$1-6$\\
\hline
$~45~$& $~60~$& $32$ &$48-53$&$6-12$\\                                                            
 \hline
 \end{tabular}\end{center}
\vskip .5cm
Finally a left double flop with ${\mathcal V}_{\rm double}$:
 \vskip 1cm
\begin{center}
\begin{tabular}{|c|c|c|c|c|}
\hline
$~{\eta^\circ_\odot} ~$& $~{\eta^\circ_\oplus} ~$&${\theta^\circ_\odot}$&${\theta^\circ_\oplus}$ &${\theta^\circ_{13}}$ \\
 \hline \hline          
$~45~$& $~30~$  &$ \sim 34$&$ 40-46$&$5-8$\\
\hline
\end{tabular}\end{center}
We see that double flops can produce a larger CHOOZ angle. Also a left flop from a family-symmetric Yukawa, 

$${\mathcal U}^{}_{MNS}={\begin{pmatrix}1&{\lambda}&{\lambda^3}\cr
{\lambda}&
1&{\lambda^2}\cr {\lambda^3}& {\lambda^2}&1\end{pmatrix}}
{\begin{pmatrix}\cos\eta^{}_\odot&\sin\eta^{}_\odot&0\cr
-\cos\eta^{}_\oplus~\sin\eta^{}_\odot&\cos\eta^{}_\oplus~\cos\eta^{}_\odot&\sin\eta^{}_\oplus\cr
\sin\eta^{}_\oplus~\sin\eta^{}_\odot&-\sin\eta^{}_\oplus~\cos\eta^{}_\odot&\cos\eta^{}_\oplus\end{pmatrix}}\ ,
$$
yields  $\eta_\odot~\sim~ 40^\circ\ ;\quad \eta_\oplus~ \sim~ 45^\circ\ ; \quad \theta^{}_{13}~\sim~0.7\,{\lambda}~\sim~9^\circ $, which we have already seen. Finally we note that  CP-violation effects can be much larger than in the quark sector. This is because the CP-violating lepton invariant\cite{JARLSKOG,GREENBERG} is 

$$J~\sim~(\lambda-\lambda^3)\,\sin\delta\ ,$$
to be compared with that in the quark sector which is of order $\lambda^6$. If the limit of zero Cabibbo mixing is meaningful for theory, analyses of the type we have just presented will assume some importance. One important remark emerges: precision measurements of the MNS matrix is quite important for theory.

\section{Conclusions}
We are beginning to read the new lepton data, but there is much work to do before a credible theory of flavor is proposed. The Seesaw Mechanism links static neutrino to physics that can never be reached by accelerators, creating a new era of the physics which centers around right-handed neutrinos. With no electroweak quantum numbers, they could hold the key to the flavor puzzles. The second large neutrino mixing angle suggests that hierarchy is independent of electroweak breaking, and occurs at grand-unified scales. 

I would like to express my gratitude to P. Bin\'etruy, M. Cribier, J. Orloff, S. Lavignac and D. Vignaud for organizing this conference {\it tr\`es sympathique} in the heart of Paris, at the Institut Henri Poincar\'e. I also wish to thank my collaborators A. Datta and L. Everett for many useful insights.

\end{document}